\documentclass[twocolumn,preprintnumbers,superscriptaddress,nofootinbib,aps,prd,floatfix]{revtex4}
\pdfoutput=1
\usepackage{enumerate}
\usepackage{amsmath,amssymb}
\usepackage{graphicx}
\usepackage{slashed}
\usepackage{float}
\usepackage{comment}
\usepackage[dvipsnames]{xcolor}
\usepackage[normalem]{ulem}
\usepackage{subfigure,orcidlink}
\usepackage{multirow,array}
\usepackage[normalem]{ulem}
\usepackage{hyperref}

\usepackage{multirow}
\usepackage{booktabs}

\begin{document}

\title{Dark Phoenix: dark matter relic from its own decay}
	
	\author{Debasish Borah \orcidlink{https://orcid.org/0000-0001-8375-282X}}
\email[Contact author: ]{dborah@iitg.ac.in}
\affiliation{Department of Physics, Indian Institute of Technology Guwahati, Assam 781039, India}
    
    \author{Satyabrata Mahapatra \orcidlink{https://orcid.org/0000-0002-4000-5071}}
	\email[Contact author: ]{satyabrata@iitgoa.ac.in}
    \affiliation{School of Physical Sciences, Indian Institute of Technology Goa, Ponda 403401, Goa, India }
    
	\author{Indrajit Saha \orcidlink{https://orcid.org/0000-0002-7459-0838}}
    \email[Contact author: ]{s.indrajit@iitg.ac.in}
    \affiliation{Department of Physics, Indian Institute of Technology Guwahati, Assam 781039, India}
    
	\author{Narendra Sahu \orcidlink{https://orcid.org/0000-0002-9675-0484}}
\email[Contact author: ]{nsahu@phy.iith.ac.in}
\affiliation{Department of Physics, Indian Institute of Technology Hyderabad, Kandi, Telangana 502285, India}
	
	\author{Vicky Singh Thounaojam \orcidlink{https://orcid.org/0009-0001-6257-5171}}
    \email[Contact author: ]{ph22resch01004@iith.ac.in}
	\affiliation{Department of Physics, Indian Institute of Technology Hyderabad, Kandi, Telangana 502285, India}
    
	\begin{abstract}
We propose a novel mechanism for generating correct relic of dark matter (DM) which otherwise gets thermally overproduced from the conventional freeze-out mechanism. The mechanism, dubbed as {\it Dark Phoenix}, relies on a transient decay window for DM after its freeze-out which brings its relic within observed limits. Due to finite-temperature effects on masses, DM $\psi$ becomes heavier than its dark sector partner $\phi$ in this window allowing it to decay. The decay of DM then stops after $\phi$ gets a sudden jump in its mass from a first-order phase transition (FOPT) driven by another scalar $\eta$. The dark sector partner $\phi$, assumed to be a charged scalar, undergoes sufficient pair annihilation during this epoch such that its late decay into DM does not overproduce the latter again. While direct-detection rates of DM remains suppressed due to small couplings, the charged scalar $\phi$ can have interesting signatures like long-lived charged track at colliders. The associated FOPT can also lead to observable gravitational waves at future experiments like LISA.

	\end{abstract}	
	\maketitle
	
\noindent
{\it Introduction}: As suggested by numerous observations in astrophysics and cosmology, approximately one-fourth of the present Universe's energy density is constituted by a non-luminous, non-baryonic form of matter, popularly referred to as the dark matter (DM) \cite{Planck:2018vyg, ParticleDataGroup:2024cfk, Cirelli:2024ssz}. Assuming DM to be of particle origin, it is well known that none of the standard model (SM) particles satisfy the criteria for being a particle DM. This has motivated several beyond standard model (BSM) studies out of which the weakly interacting massive particle (WIMP) \cite{Kolb:1990vq, Jungman:1995df, Bertone:2004pz} remains the most popular one. In such a framework, DM can be thermally produced in the early Universe due to its sizeable non-gravitational interactions with the SM bath. Subsequently, DM relic is set by its thermal freeze-out. The same DM-SM interactions can also lead to observable DM-nucleon scattering at terrestrial detectors. However, null results at direct-detection experiments like \texttt{LUX-ZEPLIN (LZ)} ~\cite{LZ:2024zvo}, \texttt{XENONnT} \cite{XENON:2025vwd}, \texttt{PandaX-4T} \cite{PandaX:2024qfu} have already ruled out a large part of the parameter space of the simplest WIMP models. This has also motivated the study of light thermal DM with mass $(m_{\rm DM} \lesssim \mathcal{O}(10 \, \rm GeV))$ where direct-detection bounds are relatively weaker. However, for DM interactions typically in the WIMP ballpark, the requirement of DM not overclosing the universe leads to a lower bound on its mass, around a few GeV \cite{Lee:1977ua, Kolb:1985nn}(with some exceptions for scalar DM \cite{Boehm:2003hm}). In the presence of light mediators between DM and SM sectors, however, one can achieve the correct relic abundance as pointed out in several works \cite{Pospelov:2007mp, DAgnolo:2015ujb, Berlin:2017ftj, DAgnolo:2020mpt, Herms:2022nhd, Jaramillo:2022mos, Borah:2024yow, Borah:2025wcc}\footnote{See also Refs.~\cite{Spergel:1999mh,Tulin:2017ara,Borah:2023sal, Borah:2022ask, Borah:2021yek, Borah:2021pet, Borah:2021rbx, Borah:2021qmi,Dutta:2022knf} where a large annihilation cross-section is achieved due to a light mediator introduced to explain DM self-interactions.}. However, such light DM with a large annihilation rate to SM often faces tight constraints from cosmic microwave background (CMB) observations \cite{Madhavacheril:2013cna, Slatyer:2015jla, Planck:2018vyg}. Such constraints can be evaded if DM is kept in the kinematically forbidden regime \cite{DAgnolo:2015ujb, DAgnolo:2020mpt, Griest:1990kh}. Instead of enhancing the annihilation cross-section, one can also consider late entropy dilution to bring the relic of thermally overproduced light DM within observed limits \cite{Evans:2019jcs, Borah:2022byb, Borah:2023sbc}.

\begin{figure}[h]
    \centering    \includegraphics[width=\linewidth]{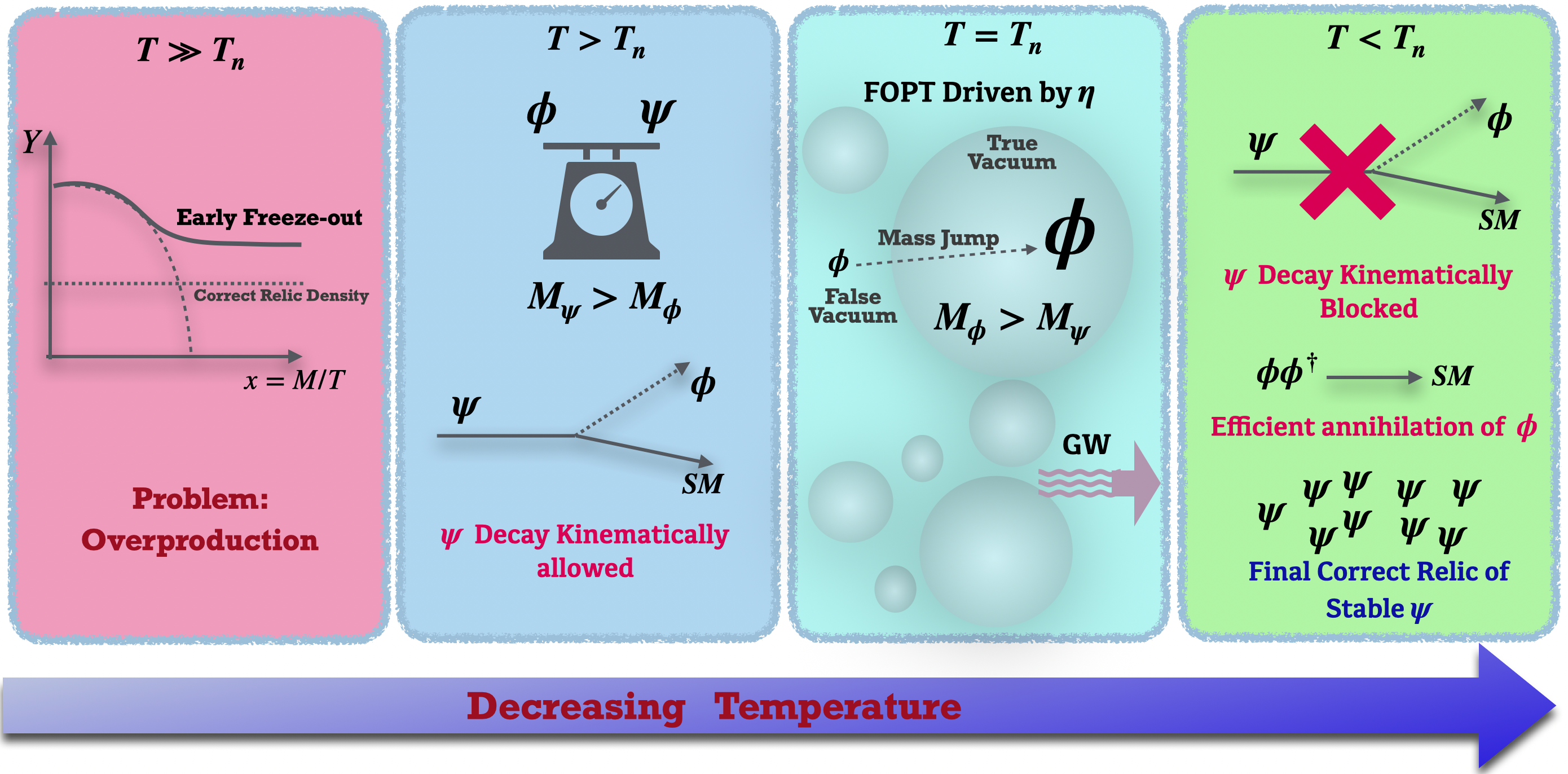}
    \caption{Schematic of the proposed mechanism.}
    \label{fig:schematic}
\end{figure}

Motivated by these, we propose a novel mechanism to generate correct relic of thermally overproduced DM by introducing a transient phase of dark matter decay. At high temperature, DM undergoes thermal freeze-out leaving an overproduced relic due to insufficient annihilation rate. This is followed by an epoch of finite duration when DM becomes the next-to-lightest stable particle (NLSP) in the dark sector due to finite-temperature effects on the masses. This transient phase of DM decay ends abruptly due to a first-order phase transition (FOPT) which again converts DM into the lightest stable particle (LSP) in the dark sector. This era of transient decay brings the relic of thermally overproduced DM within observed limits. The NLSP is assumed to have sizeable interactions with the SM bath such that it self-annihilates sufficiently during this transient era. This ensures that the decay of NLSP to DM at the end of the transient era does not lead to DM overproduction. We coin this mechanism is {\it Dark Phoenix} (DP) where final relic of DM is set by its own transient decay, as summarised in the form of a schematic in Fig. \ref{fig:schematic}. While our proposed mechanism is viable for the entire thermal DM mass range upto the unitarity upper limit on DM mass $m_{\rm DM} \lesssim \mathcal{O}(100\, \rm TeV)$ \cite{Griest:1989wd}, we focus on the light DM window with interesting detection prospects across different frontiers. It is worth mentioning that our proposed mechanism is complementary to the conversion-driven freeze-out \cite{Garny:2017rxs, DAgnolo:2017dbv, Heisig:2024mwr} where DM relic is generated due to conversion via scattering while keeping the DM stable throughout. While the usual conversion-driven freeze-out (CDFO) works without a FOPT, our mechanism relies on a FOPT which brings additional detection prospects\footnote{See earlier works \cite{Cohen:2008nb, Cui:2011qe, Baker:2016xzo, Baker:2017zwx, Bian:2018bxr, Baker:2018vos, Bian:2018mkl, Heurtier:2019beu, Baker:2019ndr, Croon:2020ntf, Elor:2021swj, Hashino:2021dvx, Bian:2021dmp, Adhikary:2024btd, Allahverdi:2024ofe, Borah:2025ydc,Mahapatra:2026fyv} connecting DM relic to FOPT.}. While transient decay of DM was also utilized in \cite{Baker:2016xzo}, it involved the breaking of the DM stabilizing symmetry followed by its restoration requiring a two-step phase transition. Our mechanism neither requires breaking of the stabilising symmetry nor a two-step phase transition. It also relies on an NLSP having sizeable interactions with the SM which brings interesting detection prospects as we discuss below. \\

\noindent
{\it The Framework}: We consider a minimal framework to implement our idea. The dark sector is composed of a singlet Dirac fermion $\psi$ and a complex scalar $\phi$ of hypercharge $1$, both of which are odd under an unbroken $Z_2$ symmetry ensuring the stability of the LSP. This leads to an interaction Lagrangian of the type
\begin{equation}
    \mathcal{L} \supset  -y \phi  \overline{\psi}e_{R} + \text{h.c.}
    \label{eq:lagrangian1}
\end{equation}
with $e_R$ being the right-handed charged lepton. Another complex singlet scalar field $\eta$ is included which drives a first-order phase transition and leads to a discontinuous change in the mass of $\phi$ preventing DM decay at lower temperatures. To realise a FOPT, we include an additional real scalar field, denoted as $\phi_2$. The relevant part of the scalar potential involving $\phi, \eta, \phi_2$ is given by
\begin{equation}
    \begin{split}
        V \supset &-\mu_{\eta}^2\eta^\dagger \eta+\lambda_{\eta}(\eta^\dagger \eta)^2+\lambda_{12}(\eta^\dagger \eta)(\phi^\dagger \phi)\\
        & +\mu_{\phi}^2\phi^\dagger \phi+\lambda_{\phi}(\phi^\dagger \phi)^2 +\frac{\lambda_{\eta\phi_2}}{2}(\phi_2 \phi_2)(\eta^\dagger \eta) \\
        & + \frac{\mu^2_{\phi_2}}{2} \phi^2_2 + \frac{\lambda_{\phi_2}}{4} \phi^4_2.
    \end{split}
\end{equation}
Due to the choice of the sign $\mu_{\eta}>0$, the field $\eta$ develops a non-zero vacuum expectation value (VEV) $v_\eta$. In order to study the finite-temperature behavior of the potential, one also has to include the Coleman-Weinberg correction $V_{\rm CW}$\cite{Coleman:1973jx} together with the thermal correction $V_{\rm th}$ \cite{Dolan:1973qd,Quiros:1999jp}. We demand the phase transition to be of first order and identify a few benchmark points consistent with this criteria, as shown in Table \ref{tab:BP}.  The same table also shows other key parameters, namely, the latent heat released $(\alpha)$, inverse duration of the transition $(\beta/\mathcal{H})$, critical temperature $(T_c)$ and nucleation temperature $(T_n)$ for the chosen benchmark points. Throughout the entire analysis, we maintain fixed values of $\lambda_\eta$ and $\lambda_{\phi_2}$, set at 0.01.

\begin{figure*}
    \centering
    \includegraphics[width=0.47\linewidth]{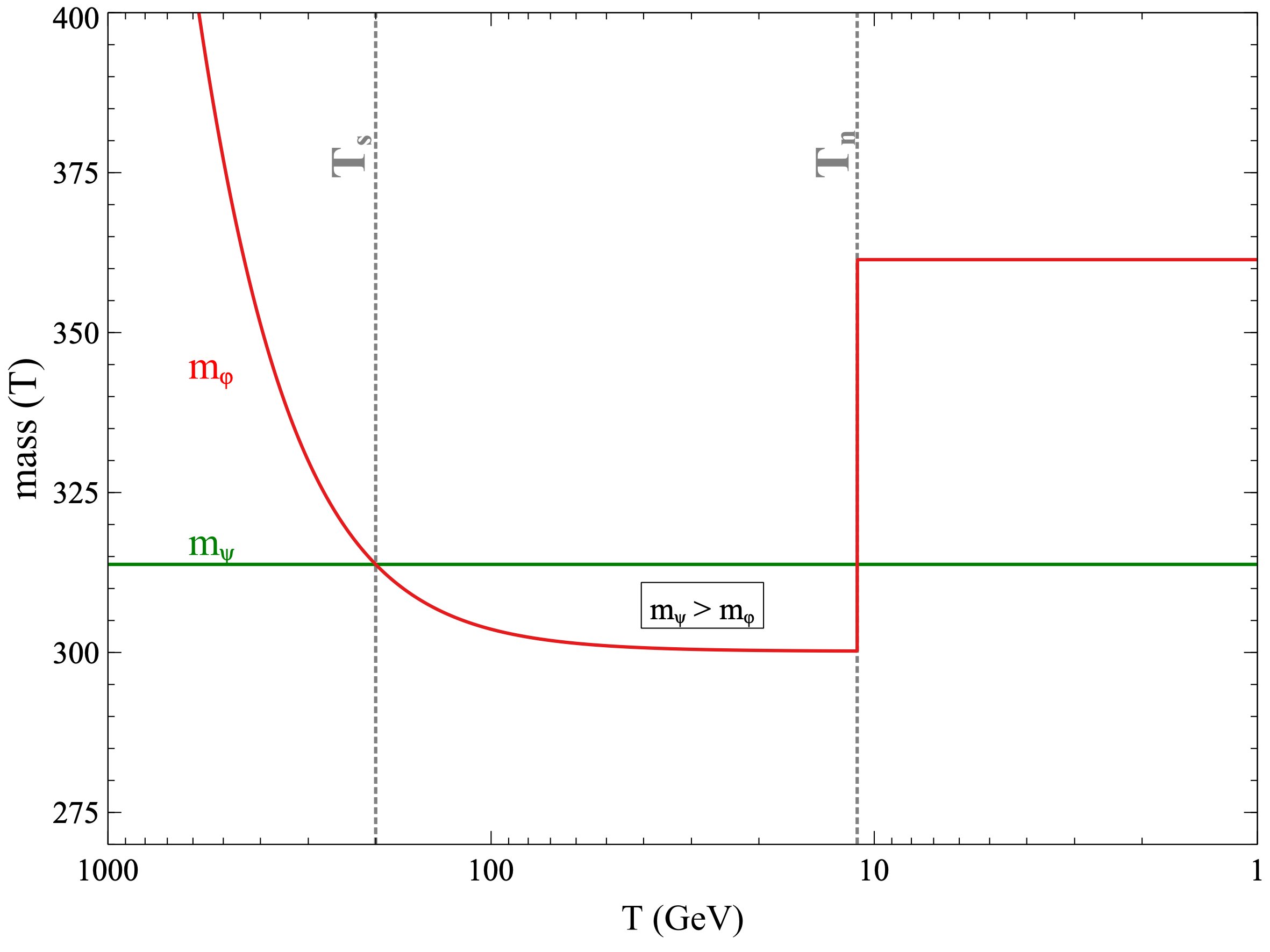}
    \includegraphics[width=0.47\linewidth]{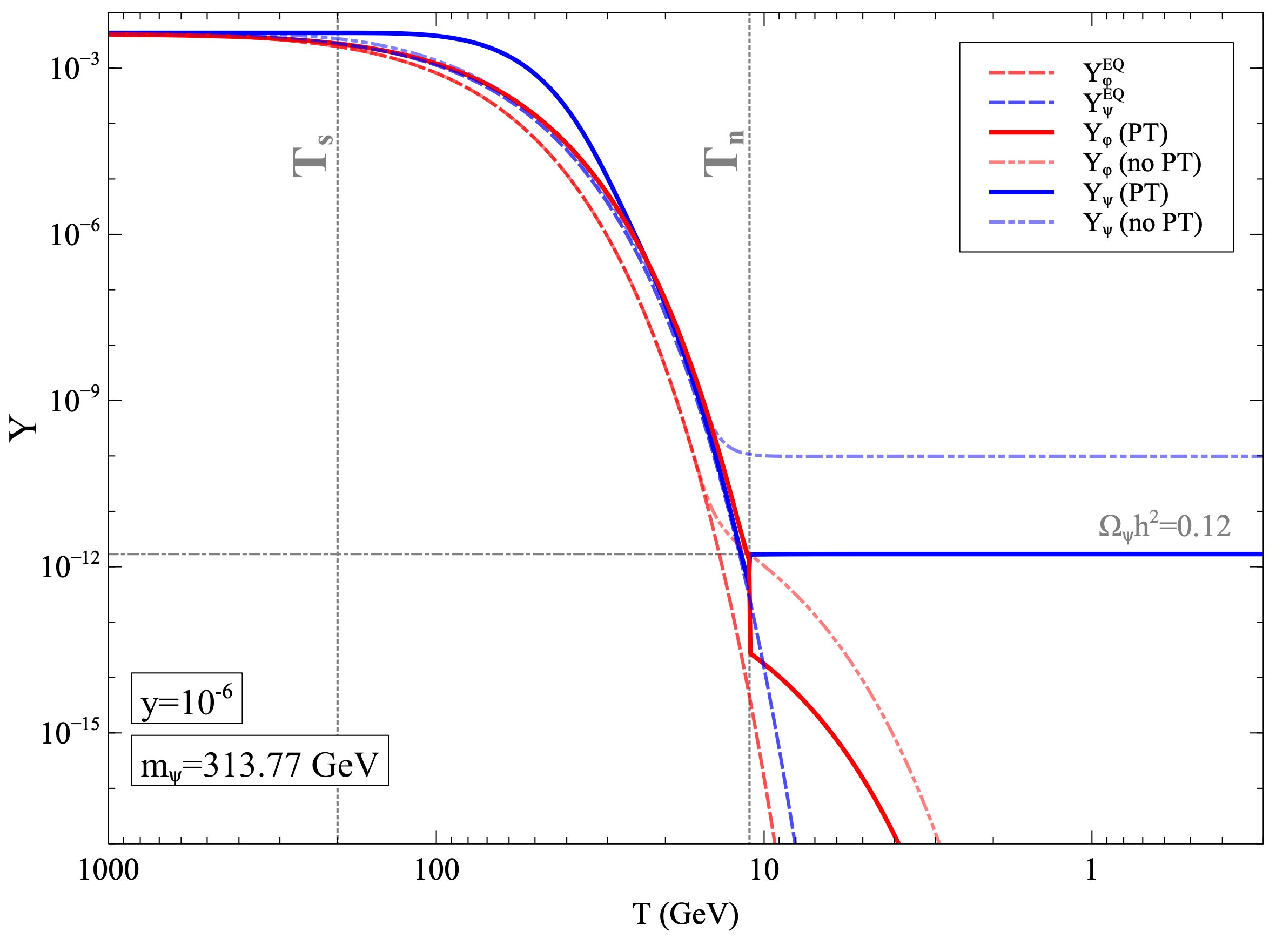}
    \caption{Left panel: Variation of the $\phi$ mass with $T$ against the fixed $\psi$ mass for BP II given in Table \ref{tab:BP}. Right panel: Evolution plot of the $\phi$ and $\psi$ particles for masses {$m_\phi = 361.4~\rm GeV, \Delta m = 47.63~\rm GeV$}. The blue curves show the abundance of $\psi$, while the red curves represent the abundance of $\phi$. The dot–dot–dashed lines correspond to the evolution assuming CDFO. The solid lines show the evolution when the $\phi$ mass arises from the FOPT with $ T_n = 10.96~\rm GeV$.}
    \label{fig:mass_window1}
\end{figure*}

In the FOPT calculations, we use the field dependent masses of $\phi, \eta, \phi_2$ as
\begin{equation}
    \begin{split}
        m_\phi^2&=\mu_{\phi}^2+\frac{\lambda_{12}}{2} \eta^2, \,\, 
        m_\eta^2 =-\mu_{\eta}^2 + 3\lambda_\eta \eta^2, \,\,m_{\phi_2}^2=\frac{\lambda_{\eta\phi_2}}{2}\eta^2,
    \end{split}
    \label{eq:field_dependent_mass}
\end{equation}
where we have assumed $\mu^2_{\phi_2}=0$ for simplicity. The corresponding temperature-dependent masses are given by 
\begin{equation}
    \begin{split}
        m_\phi^2(T)&=T^2\left(\frac{\lambda_\phi}{3} +\frac{\lambda_{12}}{12} \right), \\
        m_\eta^2(T) & =T^2\left(\frac{\lambda_\eta}{3} +\frac{\lambda_{12}}{12} +\frac{\lambda_{\eta\phi_2}}{24} \right),\\
        m_{\phi_2}^2(T) & =T^2\left(\frac{\lambda_{\phi_2}}{4} +\frac{\lambda_{\eta\phi_2}}{12} \right).
    \end{split}
    \label{eq:T_dependent_mass}
\end{equation}
The left panel of Fig. \ref{fig:mass_window1} shows the temperature evolution of the physical masses of DM $(\psi)$ and its dark sector partner $\phi$ for one of the benchmark points (BP II) given in Table \ref{tab:BP}. This clearly shows a temperature window $ T_n < T < T_s$ during which $m_\psi (T) > m_\phi (T)$ allowing transient decay of DM. In order to realise our mechanism it is important to ensure that DM freeze-out does not occur at $T< T_n$. While DM can freeze out in this window $ T_n < T <  T_s$ too, we consider the freeze-out to occur at $T_{\rm F.O.} \gg  T_s$ to show the maximal impact of our mechanism on the DM parameter space. This leads to the following constraint 
\begin{equation}
    \begin{split}
        T_{\rm F.O.} &\gg T_s \implies
        \frac{m_\psi}{x_{\rm F.O.}} \gg \sqrt{\frac{12}{\lambda_{12}}(m_\psi^2-\mu_\phi^2)}\\
        \implies \lambda_{12} &\gg 12 \times x_{\rm F.O.}^2
    \end{split}
\end{equation}
where, in the second line, we assume that $m_\psi > \mu_\phi$ and $\lambda_{12} \gg \lambda_\phi$. For the conventional freeze-out scenario driven by Boltzmann suppression, with $x_{\rm F.O.} \approx 5$–$20$, the required value of $\lambda_{12}$ becomes non-perturbative. Consequently, we are led to consider an ultra-relativistic freeze-out scenario \cite{Henrich:2025sli}, where the freeze-out temperature satisfies $T_{\rm F.O.} \gg m_\psi$. In order to ensure that the thermally overproduced DM can decay for a sufficient duration during $T_n < T <  T_s$, we also get an upper bound on the nucleation temperature $ T_n \lesssim 0.03~m_{\psi}$ the details of which are given in Appendix \ref{app:Tn}.
\\

\begin{table*}
    \centering
    \caption{BP comparisons. All dimensionful parameters are expressed in GeV.}
    \begin{tabular}{|c|c|c|c|c|c|c|c|c|c|c|c|c|c|}
        \hline
        \textbf{BP} & $\mu_\phi$ & $\lambda_\phi$ & $\lambda_{12}$ & $v_\eta$ & $m_\eta$ & $m_\phi(T=0)$ & $m_{\phi_2}$ & $\lambda_{\eta \phi_2}$& $T_c$ & $T_n$ & $\alpha$ & $ \beta/\mathcal{H}$\\
        \hline\hline
        \textbf{I} & 272.18 & 0.01 & 0.6 & 120 & 16.97 & 280.0 & 131.64 & 2.4 & 34.36 & 10.26 & 0.74 & 82.60 \\
        \hline
        \textbf{II} & 300.20 & 0.01 & 2.5 & 180 & 25.45 & 361.4 & 182.19 & 2.0 & 64.35 & 10.96 & 2.08 & 4.51 \\
        \hline
        \textbf{III} & 749.25 & 0.01 & 2.9 & 550 & 77.78 & 1000.0 & 458.68 & 1.4 & 196.56 & 26.48 & 6.33 & 33.86 \\
        \hline
    \end{tabular}
    \label{tab:BP}
\end{table*}

\begin{figure*}
    \centering
    \includegraphics[width=0.47\linewidth]{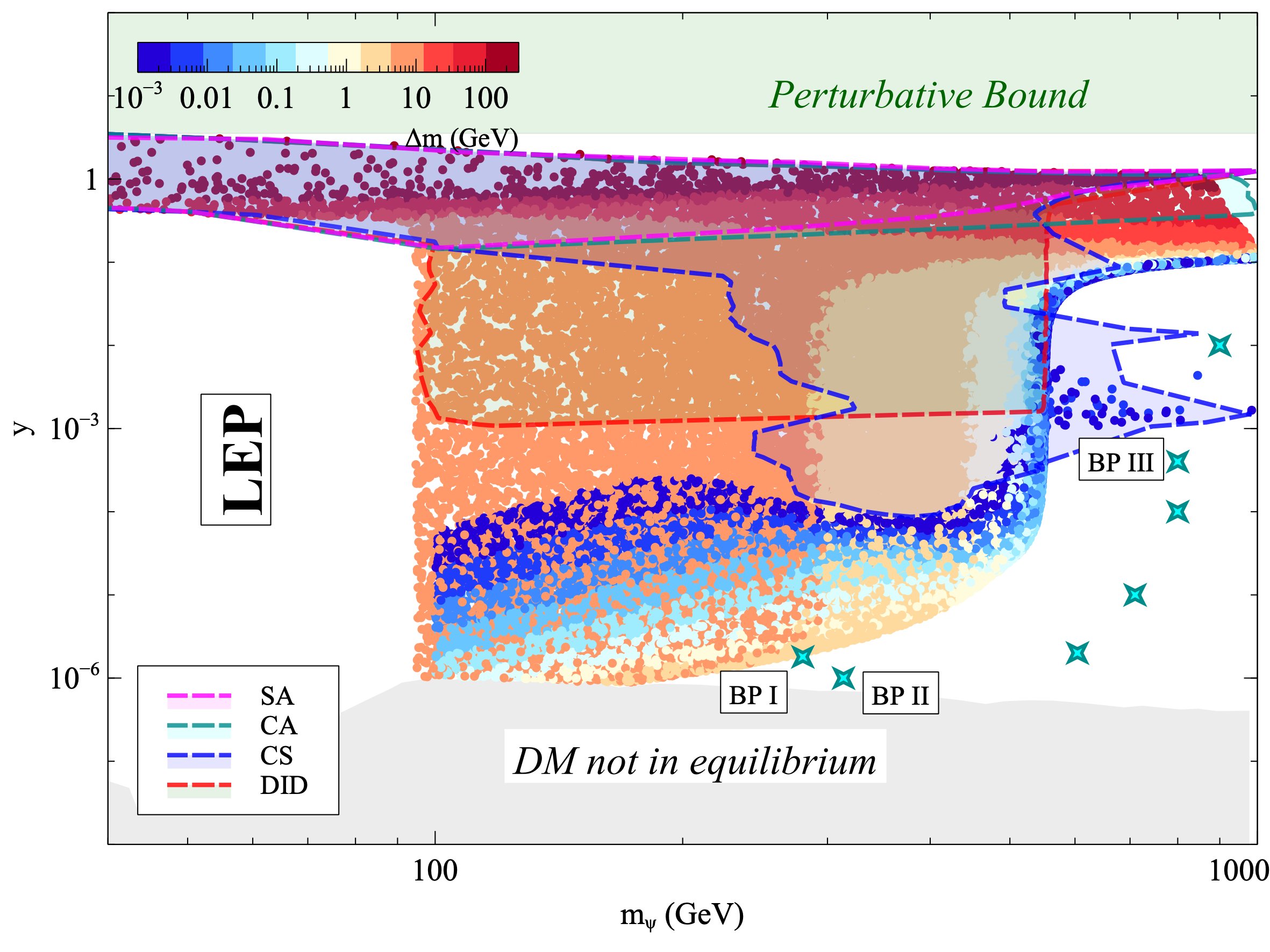}
    \includegraphics[width=0.47\linewidth]{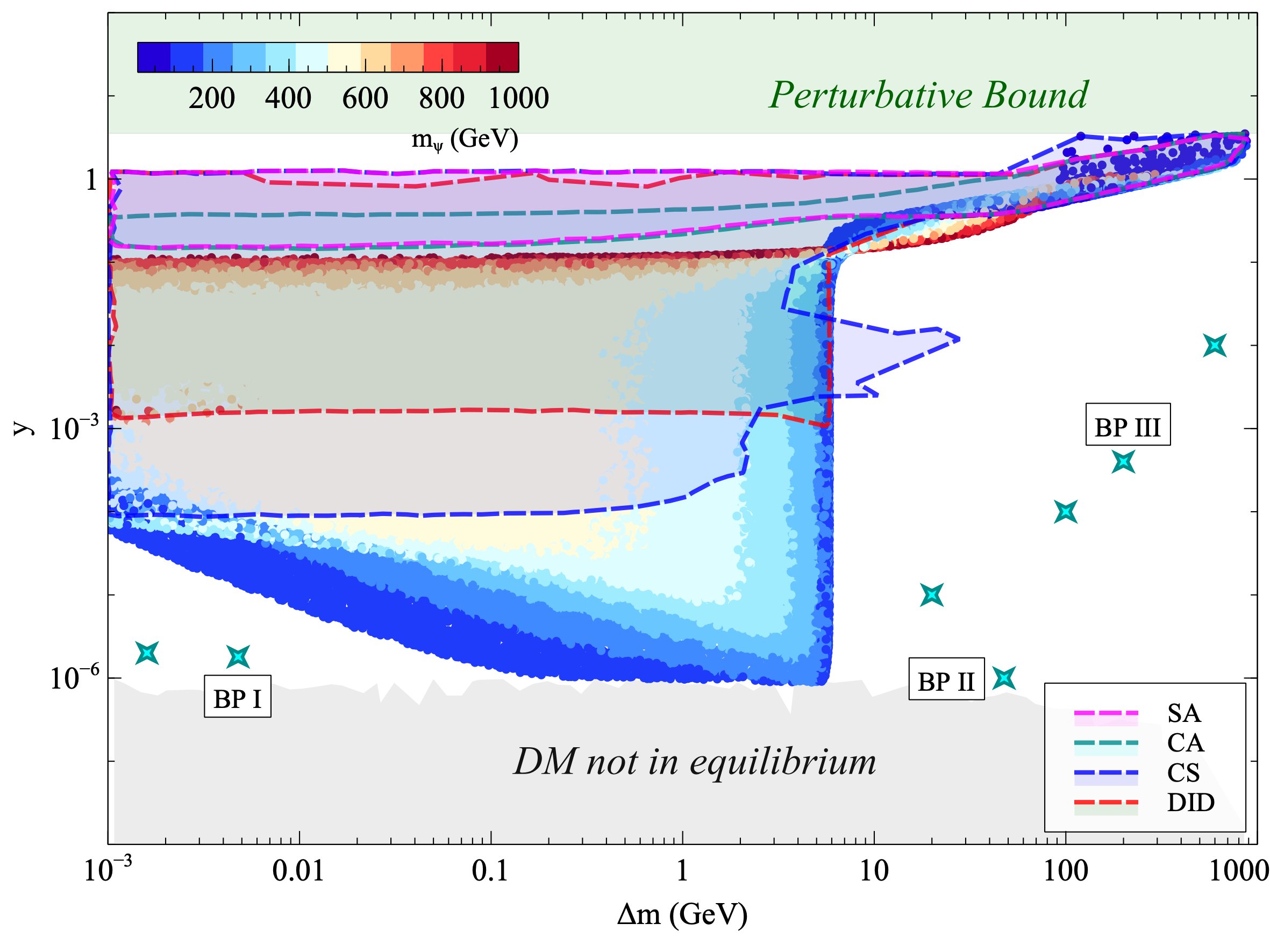}
    \caption{The parameter space consistent with the observed relic density ($\Omega {\rm h}^2=0.12 \pm 0.001$) is shown in the $y-m_\psi$ plane (left panel) and the $y-\Delta m$ plane (right panel). The different contours indicate the dominant mechanisms responsible for setting the relic abundance: self-annihilations (SA, magenta), co-annihilations (CA, green), co-scattering (CS, blue), and decay with inverse decay (DID, red). In the left panel, the white region on the left side is excluded by the LEP lower bound on the mass of $\phi$. In both the panels, the white region on the far right leads to thermal overabundance in the absence of the dark phoenix mechanism. The diamond-shaped cyan points denote a selection of viable points for which the DM relic can be generated through the mechanism proposed in this work. We explicitly highlight three of these as specific benchmark points (BPs), with their full details provided in Table~\ref{tab:BP}.
    }
    \label{fig:relic_summary}
\end{figure*}

\noindent
{\it Results and Discussion}: The right panel of Fig. \ref{fig:mass_window1} illustrates how the comoving number densities of DM and $\phi$ evolve, highlighting the effect of transient DM decay aided by the FOPT. This figure corresponds to benchmark point BP II listed in Table \ref{tab:BP}. When neither FOPT nor transient decay is present, DM first decouples relativistically and then proceeds through standard CDFO. Given the chosen model parameters, ordinary CDFO alone cannot bring the DM relic density down to observed levels, as indicated by the dot-dot-dashed blue line. In contrast, our dark phoenix mechanism uses the transient decay window to successfully reduce the DM relic abundance to the observed value, shown by the solid blue line. Although DM decays into $\phi$ during this window, the resulting $\phi$ population does not accumulate substantially, thanks to its efficient depletion through gauge and scalar portal interactions. Because $\phi$ possesses non-zero hypercharge, it interacts with the electroweak neutral gauge bosons. It also couples to the scalar field $\eta$ with sizeable quartic coupling. Both of these interactions keep $\phi$ thermally coupled to the bath. The gauge coupling of $\phi$ is fixed by its hypercharge, but the portal coupling between $\eta$ and $\phi$ i.e. $\lambda_{12}$, can be adjusted independently. Given that $\eta$ itself undergoes FOPT with a preferably large $\lambda_{12}$ coupling, and that its mass is considerably smaller than that of $\phi$ (see Table~\ref{tab:BP}), this opens up an extra annihilation channel for $\phi$. This ensures that $\phi$'s overall annihilation rate into SM particles remains sufficiently high such that any late decay of $\phi$ occurring outside the transient window does not push the DM abundance above the observed relic density. 


Fig.~\ref{fig:relic_summary} summarizes the viable parameter space for DM relic density, covering mechanisms including the standard annihilation (SA), co-annihilation (CA) \cite{Griest:1990kh}, co-scattering (CS) \cite{DAgnolo:2017dbv}, and decay and inverse decay (DID) \cite{Frumkin:2021zng}. For this numerical scan, the portal coupling $\lambda_{12}$ is set to $2$ to ensure efficient annihilation of $\phi$. Taken together, these channels account for the full region of parameter space consistent with the observed relic density, as shown in both panels of Fig.~\ref{fig:relic_summary}. The left panel presents the allowed parameter space in the $y$--$m_\psi$ plane, and the right panel shows the corresponding region in the $y$--$\Delta m$ plane. The gray shaded region denotes where DM fails to thermalize in the early Universe, and the light-green shaded region marks the perturbative breakdown. The cyan-colored $\star$-shaped points correspond to the parameter space where CDFO alone overproduces DM, but where the correct relic density is recovered through the proposed dark phoenix mechanism. Although the Yukawa coupling $y$ must be sufficiently large to ensure thermalization, all points in Fig.~\ref{fig:relic_summary} are consistent with direct detection constraints, the details of which are given in Appendix~\ref{app:int_rate}. 

\begin{figure}
    \centering
    \includegraphics[width=0.95\linewidth]{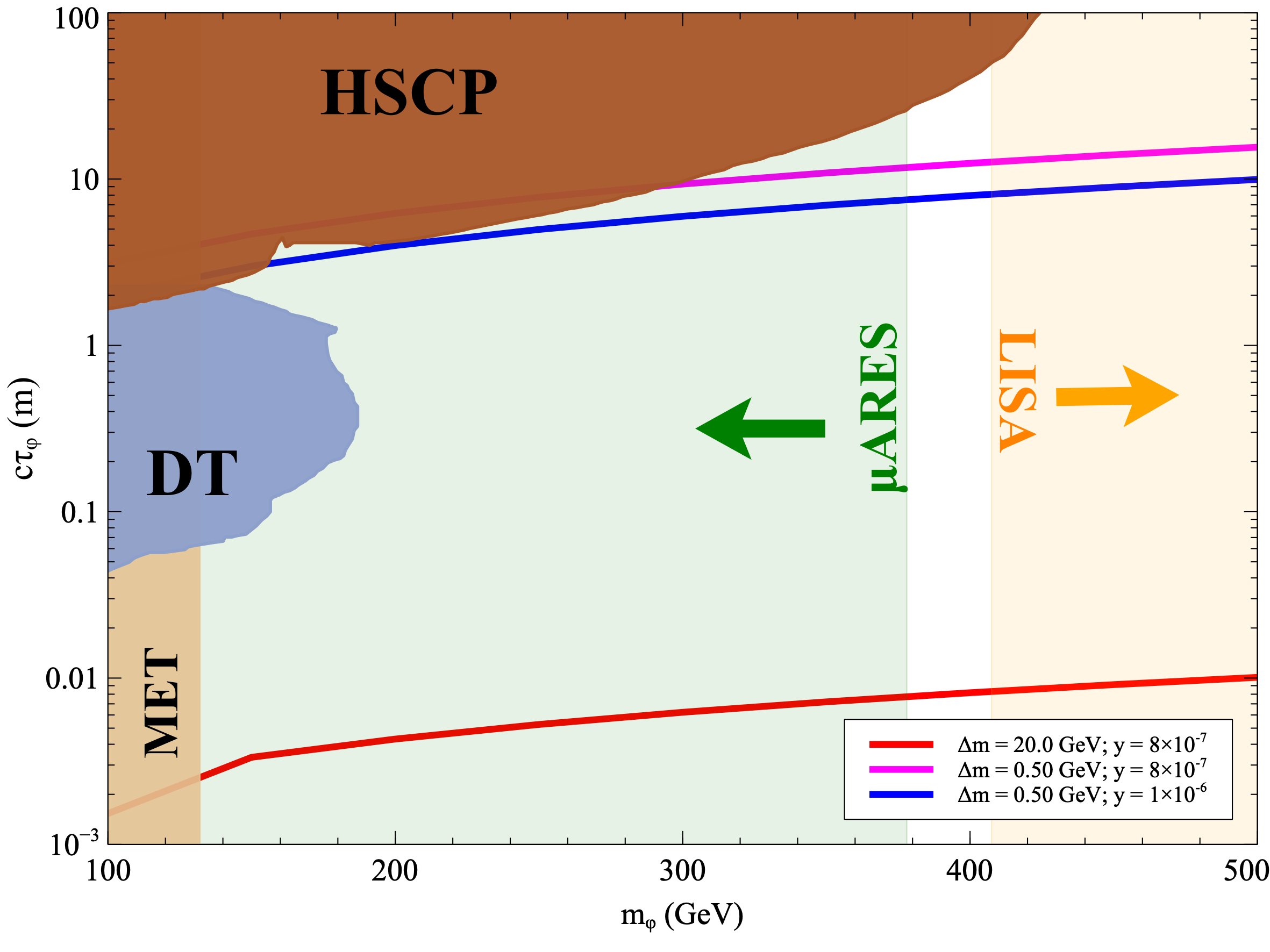}
    \caption{The decay length is shown together with the 95\% CL exclusion limits from the HSCP, DT, and MET searches. Three benchmark points are presented to illustrate the impact of $\Delta m$ and the Yukawa coupling $y$. The red ($\Delta m=20~\rm GeV$) and magenta ($\Delta m=0.5~\rm GeV$) solid curves correspond to the same Yukawa coupling, $y=8 \times 10^{-7}$, highlighting the effect of varying $\Delta m$. In contrast, the blue ($y=10^{-6}$) and magenta ($y=8\times10^{-7}$) solid curves correspond to the same mass splitting, $\Delta m=0.5~\rm GeV$, illustrating the dependence on the Yukawa coupling.}
    \label{fig:displaced_vertex}
\end{figure}

Owing to its gauge charge, $\phi$ can be pair-produced copiously at colliders, subsequently decaying into DM and a charged lepton $e_R$, with the decay width
\begin{equation}
    \Gamma_{\phi}=\frac{y^2 m_\phi}{32 \pi} \left(1-\frac{m_\psi^2}{m_\phi^2} \right)^2.
    \label{eq:phi_decay width}
\end{equation}
For sizeable Yukawa coupling $y$, such decays into a charged lepton plus missing transverse energy (MET) are constrained by LHC searches for prompt signatures \cite{ATLAS:2019lng, CMS:2024gyw}. However, since reproducing the correct DM relic density requires a relatively small $y$, both for ultra-relativistic freeze-out and for the regulated transient decay, $\phi$ can instead give rise to striking collider signatures such as disappearing tracks (DT) \cite{ATLAS:2017oal, CMS:2020atg} or heavy stable charged particles (HSCP) \cite{CMS:2013czn, ATLAS:2019gqq}, both of which are subject to dedicated LHC searches. These constraints, adapted from \cite{Heisig:2024mwr}, are shown as shaded regions in the left panel of Fig.~\ref{fig:displaced_vertex}. The scalar mass is taken above 100 GeV in accordance with the LEP bound \cite{lepsusy}.

The decay length of $\phi$ is shown for three benchmark points in Fig.~\ref{fig:displaced_vertex}. The solid magenta line corresponds to $\{\Delta m = 0.50~\text{GeV},\, y = 8\times10^{-7}\}$, while the solid blue and solid red lines correspond to $\{\Delta m = 0.50~\text{GeV},\, y = 10^{-6}\}$ and $\{\Delta m = 20~\text{GeV},\, y = 8\times10^{-7}\}$, respectively, illustrating the sensitivity to variations in $y$ and $\Delta m$. The nucleation temperatures required to achieve the correct relic density for all three benchmark points satisfy the previously discussed constraint, $T_n < 0.03 m_\psi$.
It is also worth noting that one of the benchmark points with $\Delta m = 0.50~\text{GeV}$ can reproduce the correct relic density even in the absence of a FOPT. For the fixed values of $\alpha = 0.1$ and $\beta/\mathcal{H} = 500$, the light green and light orange shaded regions represent the parameter space accessible to future experiments $\mu$ARES~\cite{Sesana:2019vho} and LISA~\cite{LISA:2017pwj}, respectively. The details of the GW spectra for our benchmark points shown in Table~\ref{tab:BP} are given in Appendix \ref{appen2}.

Finally, since the FOPT is required to occur below the mass scale of DM or $\phi$, and given that the unitarity upper bound on $m_\phi$ restricts efficient pair annihilation to $m_\phi \lesssim \mathcal{O}(100~\text{TeV})$ \cite{Griest:1989wd}, the predicted GW signal peaks at frequencies $\lesssim 0.1~\text{Hz}$. \\

\noindent
{\it Conclusions}: We have proposed a novel mechanism to revive the thermal dark matter parameter space where conventional freeze-out mechanism leads to overproduction. We introduce a transient decay window for DM after its freeze-out which brings its relic within observed limits and refer to this mechanism as the Dark Phoenix mechanism. Considering a toy model where DM is a Dirac singlet fermion $\psi$ coupling to a singly charged scalar $\phi$ and right-handed charged leptons, we focus on the parameter space where DM gets thermally overproduced via relativistic decoupling. While finite-temperature effects on the masses cause the physical mass of $\phi$ to decrease at lower temperature, a sudden jump in its mass at a lower temperature caused by another scalar undergoing a first-order phase transition ensures the stability of DM at lower temperatures. We calculate the relic abundance of DM by considering all possible $\psi \rightarrow \phi$ conversion processes including transient decay by solving the relevant Boltzmann equations. In addition to identifying the parameter space consistent with correct relic, we also point out the notable changes our Dark Phoenix mechanism brings compared to the conversion driven freeze-out or coscattering DM studied in the literature. The gauge interactions of $\phi$ not only ensures its sufficient pair annihilation crucial for avoiding late increase in DM density from its decay, but also enhances its production at colliders. While DM direct-detection prospects remain suppressed due to small Yukawa couplings, the charged scalar can lead to observable signatures like long-lived charged tracks in colliders. In addition, the first-order phase transition at a scale below to DM mass brings interesting correlations with GW peak frequencies, within reach of future experiments like LISA. \\

\noindent 
{\it Acknowledgements}: The work of D.B. is supported by the Science and Engineering Research Board (SERB), Government of India grant CRG/2022/000603.

\twocolumngrid
	\bibliographystyle{JHEP}
	\bibstyle{apsrev}

\providecommand{\href}[2]{#2}\begingroup\raggedright\endgroup

\appendix

\section{Bound on nucleation temperature}
\label{app:Tn}
\begin{figure}
    \centering
    \includegraphics[width=0.85\linewidth]{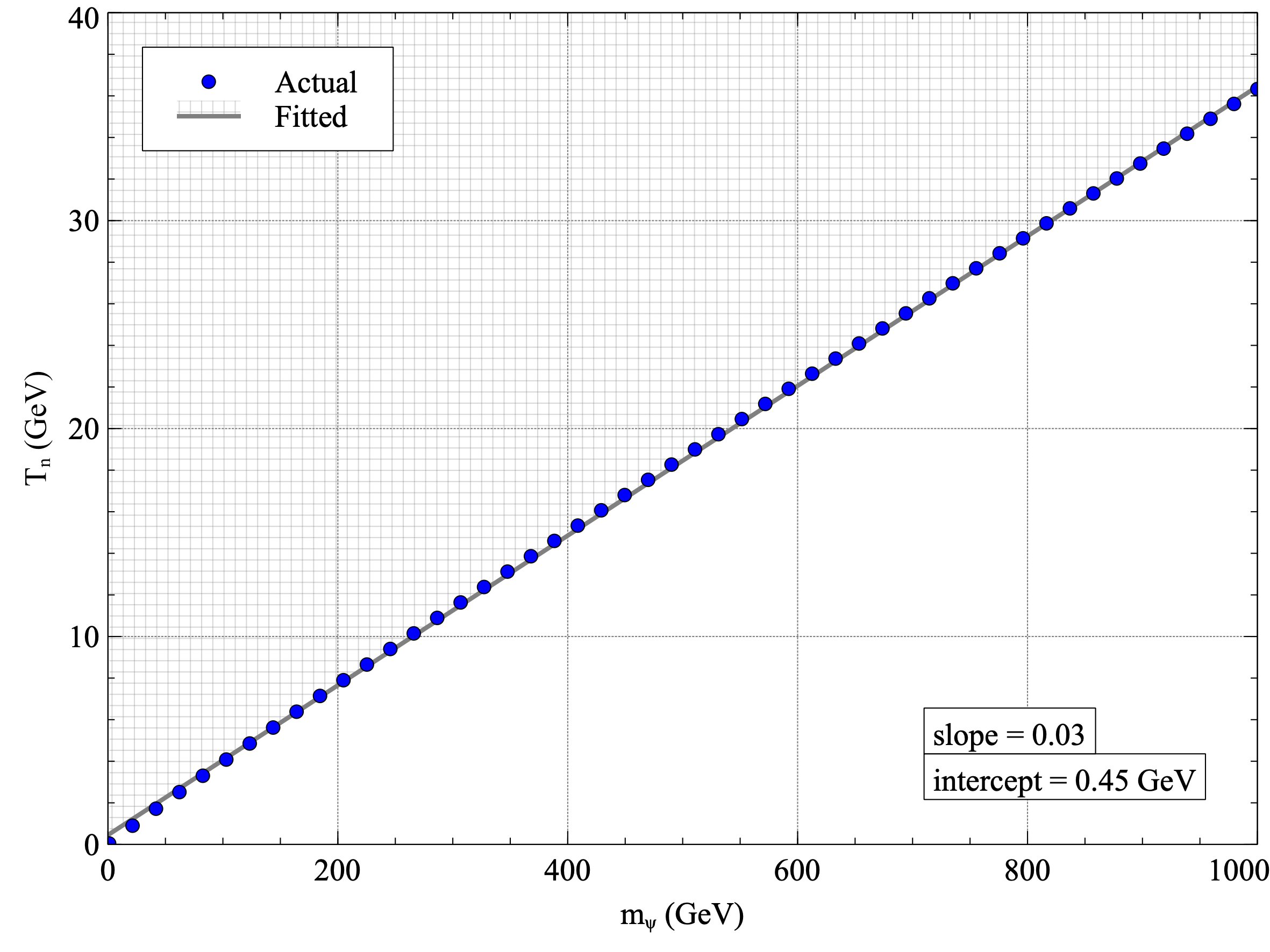}
    \caption{The figure shows the values of $T$ obtained from the condition that $Y_{\psi}^{\rm EQ}$ equals the observed abundance for a given DM mass, $m_\psi$, represented by the blue dots. The gray line corresponds to a linear fit to the data, with a slope of $0.03$ and an intercept of $0.45$ GeV. Any realization with $T_n$ above this line will result in an overproduction of the DM relic abundance (see text for further details).}
    \label{fig:Tn_mDM}
\end{figure}

An important feature of this mechanism is that the temperature window must close once the DM abundance reaches a value close to the required relic abundance. This requirement provides a rough estimate of $T_n$, which can be obtained by equating the equilibrium abundance of DM, evaluated at $T_n$, to $Y_{\rm req}^\infty$:
\begin{equation}
    \begin{split}
        \left. Y_\psi^{\rm EQ}(T)\right |_{T_n} &= Y_{\rm req}^\infty\ \\
        &=\frac{\rho_c}{m_\psi}\Omega h^2_{\rm DM}.
    \end{split}
\end{equation}
In the second line, we have used the definition of $\Omega {\rm h}^2$ to express $Y_{\rm req}^\infty$. The resulting values of $T_n$ for DM masses in the range $[1,1000]$ are shown in Fig.~\ref{fig:Tn_mDM}. We further fit the data with a straight line, obtaining a slope of $0.03$ and an intercept of $0.45$ GeV.

Therefore, if a realization of a FOPT scenario predicts a nucleation temperature larger than the value indicated in Fig.~\ref{fig:Tn_mDM} for a given DM mass, the proposed mechanism fails to reproduce the observed DM relic abundance, even when all other requirements, including a sufficiently large annihilation rate of $\phi$, are satisfied. Furthermore, increasing the Yukawa coupling does not enhance the effectiveness of the mechanism, since the same coupling also increases the repopulation of $\psi$ through $\phi$ decay outside the temperature window. Consequently, a conservative upper bound on the nucleation temperature in terms of the DM mass can be expressed as
\begin{equation}
     T_n \lesssim 0.03~m_{\psi}.
\end{equation}

\section{Details of dark matter phenomenology}
\label{app:int_rate}
\subsection{DM Relic}
Contrary to the conventional dark matter freeze-out picture where decoupling from the SM bath occurs after the particles become non-relativistic, our mechanism requires decoupling to take place while the particle is still relativistic, resulting in qualitatively distinct thermal behavior. As demonstrated in \cite{Henrich:2025sli}, such a scenario can be consistently realized.
The DM decouples while still relativistic, leaving behind a large relic abundance, a scenario commonly referred to in the literature as ultra-relativistic freeze-out (UFO).

Following this UFO phase, the DM undergoes additional processes that can further modify its number density. The Boltzmann equations governing the resulting final abundance are given by
\begin{equation}
    \begin{split}
        \frac{dY_\psi}{dT}&=\frac{1}{3\mathcal{H}(T)}\frac{ds}{dT}\left [\frac{\Gamma_{\psi \rightarrow \phi e_R}}{s} \left (Y_\psi(T) - \frac{Y_\psi^{\rm EQ}}{Y_\phi^{\rm EQ}}Y_\phi(T) \right) \right.\\
        & \quad \quad -\left. \frac{\Gamma_{\phi\psi}}{s} \left (Y_\phi(T) - \frac{Y_\phi^{\rm EQ}}{Y_\psi^{\rm EQ}}Y_\psi(T) \right)  \right]\\
        \frac{dY_\phi}{dT}&=\frac{1}{3\mathcal{H}(T)}\frac{ds}{dT} \left[\langle \sigma v \rangle \left(Y_\phi^2 -(Y_\phi^{\rm EQ})^2 \right) \right.\\
        &\quad \quad -\frac{\Gamma_{\psi \rightarrow \phi e_R}}{s} \left (Y_\psi(T) - \frac{Y_\psi^{\rm EQ}}{Y_\phi^{\rm EQ}}Y_\phi(T) \right)\\
        &\quad \quad \quad +\left. \frac{\Gamma_{\phi\psi}}{s} \left (Y_\phi(T) - \frac{Y_\phi^{\rm EQ}}{Y_\psi^{\rm EQ}}Y_\psi(T) \right)  \right]
    \end{split}
\end{equation}
where, $\Gamma_{\psi \rightarrow \phi e_R}$ is given by
\begin{equation}
    \Gamma_{\psi \rightarrow \phi e_R}=\frac{y^2 m_\psi}{32 \pi} \left(1-\frac{m_\phi(T)^2}{m_\psi^2} \right)^2
    \label{eq:phi_decay width}
\end{equation}
and $\Gamma_{\phi\psi}$ denotes the total interaction rate consisting of both the decay width of $\phi$, i.e $\Gamma_{\phi \rightarrow\psi e_R}$ and the co-scattering rate corresponding to the process $\psi e_R \leftrightarrow \phi \gamma$, where $\gamma$ denotes the SM photon. Thus
\begin{equation*}
    \Gamma_{\phi\psi}=\Gamma_{\phi \rightarrow\psi e_R}+\langle \sigma v \rangle_{\psi e_R \leftrightarrow \phi \gamma} n^{\rm eq}_{e}.
\end{equation*}
Moreover, it should be emphasized that $\Gamma_{\psi \rightarrow \phi e_R}$ is non-zero only within the specified temperature window [$T_s,T_n$] and vanishes outside it. 

The Lagrangian in Eq.~\ref{eq:lagrangian1} governs the decay width of the dark matter particle $\psi$ within this interval. However, the same Lagrangian also allows $\phi$ to decay into $\psi$ once outside the window. If this were the sole decay mode available to $\phi$, then after the temperature falls below $T_n$, the $\phi$ population would regenerate the dark matter abundance, bringing it back close to its initially large value. In that scenario, the mechanism would fail to efficiently suppress the relic density. Hence, the scalar $\phi$ must stay in equilibrium with the SM bath for as long as possible, so that its abundance is depleted through annihilations into other channels.

\subsection{Direct detection of DM}\label{app:DM_nucleon}

\begin{figure}[h]
    \centering
    \includegraphics[width=0.5\linewidth]{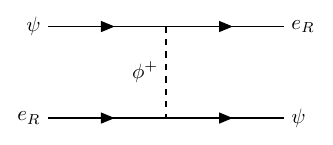}
    \caption{Feynman diagrams illustrating the process $\psi~e \leftrightarrow \psi~e$ relevant for ${\rm DM}-e$ direct detection.}
    \label{fig:dd_dm_e}
\end{figure}


In our setup, Eq.~\ref{eq:lagrangian1} mediates DM-electron scattering as shown in Fig.~\ref{fig:dd_dm_e}. A distinctive feature is that the mediator ($\phi$) carries electric charge, enforcing a collider-bound mass of $m_\phi \gtrsim 100~\rm GeV$. Consequently, the DM mass is naturally close to $m_\phi$, with a mass splitting given by $\Delta m$. The detection cross-section is given by
\begin{equation}
    \sigma_{\rm \psi-e} =\frac{y^4}{\pi} \frac{\mu_{\rm \psi e}^2}{{m^4_{\phi}}}
\end{equation}
where $\mu_{\rm \psi e}$ denotes the reduced mass of the DM-$e$ system. For $m_\psi \gg m_e$, the scattering kinematics differ from the sub-GeV case: the maximum electron recoil energy is set by $m_e$ and the DM halo velocity, resulting in keV-scale recoils. The total scattering rate saturates as the reduced mass approaches $m_e$, making the sensitivity of electron recoil searches largely dependent on the Yukawa coupling and mediator mass rather than $m_\psi$. Large-volume detectors such as XENON10/XENON100 \cite{Essig:2017kqs} are sensitive to this energy range.

Achieving signals within current or near-future sensitivity requires $y \geq 1$, accessible to experiments like xenon-based detectors. In contrast, the mechanism under consideration predicts $y \sim 10^{-6}$ ($\sigma_{\rm \psi-e}\sim 10^{-67}~\rm cm^2$ for $m_\phi=100~\rm GeV$), suppressing the scattering rate well below experimental reach. As a result, despite a tree-level coupling to electrons, the model naturally evades both current and projected constraints. This illustrates a general feature of heavy leptophilic DM: even with direct electron couplings, the combination of mediator mass and weak Yukawa coupling can render detection exceedingly challenging.

\begin{figure}[h]
    \centering
    \includegraphics[width=0.45\linewidth]{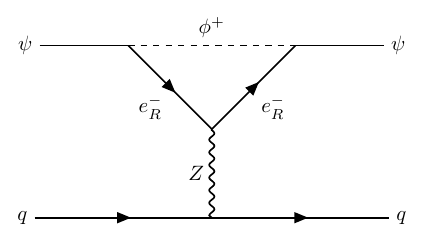}
    \includegraphics[width=0.45\linewidth]{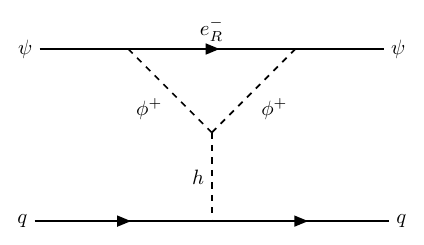}
    \caption{Feynman diagrams illustrating the spin-dependent process (left panel) and the spin-independent process (right panel) relevant for direct detection.}
    \label{fig:dd_diagram}
\end{figure}

While the Lagrangian in Eqn.~\ref{eq:lagrangian1} does not generate any tree-level DM-nucleon scattering, it can arise at the one-loop level, mediated by the $Z$ boson or the Higgs boson, corresponding to spin-dependent (SD) and spin-independent (SI) processes, respectively, as illustrated in Fig.~\ref{fig:dd_diagram}.

The effective Lagrangian for the spin-dependent and spin independent DM detection is
\begin{equation}
    \mathcal{L}_{eff} \supset
        \begin{cases}
          \xi_q \overline{\psi}\gamma^\mu \gamma^5 \psi \overline{q}\gamma_\mu \gamma^5 q, & \text{SD} \\
          \Lambda_q \overline{\psi}\psi \overline{q}q,  & \text{SI}
        \end{cases}
\end{equation}
where the effective couplings are given by
\begin{equation}
    \begin{split}
        \xi_q&=\frac{y^2a_q}{32 \pi^2 M_Z^2} \left[ (v_l+a_l)\mathcal{G}_2\left(\frac{m_{\psi}^2}{m_{\phi}^2}\right) \right]\\
    \Lambda_q&=-\frac{y^2}{16 \pi^2 m_h^2 m_{\psi}} \left[ \lambda_{h\phi} \mathcal{G}_1 \left(\frac{m_{\psi}^2}{m_{\phi}^2} \right) \right]
    \end{split}
\end{equation}
Here, $a_l=\frac{-g}{2 c_W}\frac{1}{2}$, $v_l=\frac{-g}{2 c_W}\left( \frac{1}{2}-2 s_W^2 \right)$, $a_q=\frac{1}{2} \left( -\frac{1}{2} \right)$ for $q=u,c,t(d,s,b)$, and the loop function $\mathcal{G}_1(x)$ and $\mathcal{G}_2(x)$ are given by
\begin{equation}
    \begin{split}
        \mathcal{G}_1(x)&=\frac{x+(1-x)\ln(1-x)}{x}\\
        \mathcal{G}_2(x)&=-1+\frac{2(x+(1-x){\rm ln}(1-x))}{x^2}.
    \end{split}
\end{equation}
Then, the spin-independent scattering cross section of $N_1$ off a proton is given by
\begin{equation}
    \sigma_{SI}=\frac{4}{\pi}\frac{m_{\psi}^2m_p^2}{(m_{\psi}+m_p)^2}m_p^2 \left( \frac{\Lambda_q}{m_q} \right)^2 f_p^2 ,
\end{equation}
where $m_p$ denotes the proton mass and $f_p \approx 0.3$ \cite{Hoferichter:2017olk} is the scalar form factor and the spin-dependent cross section per nucleon $N$ is given by \cite{Jungman:1995df}.
\begin{equation}
    \sigma^{SD}_{{\rm DM-N}}=\frac{16}{\pi}\frac{m_{\psi}^2m_N^2}{(m_{\psi}+m_N)^2}J_N(J_N+1)\xi_N^2,
\end{equation}
where $\xi_N=\sum_{q=u,d,s} \Delta^N_q \xi_q$ with $\Delta^N_u=0.842,\Delta^N_d=-0.427$ and $\Delta^N_s=0.085$ \cite{HERMES:2006jyl}. For $m_\phi = 278.79~\rm GeV$ and $m_\psi = m_\phi - 0.79~\rm GeV$, the spin-independent cross-section is of the order of $10^{-70}~\rm cm^2$, while the spin-dependent cross-section is of the order of $10^{-94}~\rm cm^2$. Both are far too small to be probed by current direct detection experiments or their projected future sensitivities. Therefore, direct-detection experiments do not impose additional constraints on the parameter space of our model.

\section{Gravitational waves from FOPT}
\label{appen2}

In a first-order phase transition, the system transitions from a metastable false vacuum to a stable true vacuum via nucleation of bubbles. The dynamics of bubble nucleation, expansion, and collision~\cite{Turner:1990rc,Kosowsky:1991ua,Kosowsky:1992rz,Kosowsky:1992vn,Turner:1992tz}, as well as the resulting sound waves~\cite{Hindmarsh:2013xza,Giblin:2014qia,Hindmarsh:2015qta,Hindmarsh:2017gnf} and turbulence in the plasma~\cite{Kamionkowski:1993fg,Kosowsky:2001xp,Caprini:2006jb,Gogoberidze:2007an,Caprini:2009yp,Niksa:2018ofa}, act as sources of a stochastic gravitational wave (GW) background.
\begin{align}
\Omega_{\rm GW}(f) = \Omega_\phi(f) + \Omega_{\rm sw}(f) + \Omega_{\rm turb}(f).    
\end{align}
The GW spectrum is determined by a set of phase transition parameters: the latent heat normalized to the radiation energy density $\alpha$, the inverse duration of the transition $\beta/\mathcal{H}$, the bubble wall velocity $v_w$, and the fraction of energy transferred to the plasma or the scalar field. The characteristic peak frequency $f_{\rm peak}$ and amplitude $\Omega_{\rm GW} h^2$ can be estimated using semi-analytic expressions derived from hydrodynamic simulations of expanding bubbles. The stochastic gravitational-wave background generated from bubble wall collisions can be parametrized as~\cite{Caprini:2015zlo}
\begin{eqnarray}
\Omega_\phi h^2 &=& 1.67 \times 10^{-5}
\left(\frac{100}{g_*}\right)^{1/3}
\left(\frac{\mathcal{H}}{\beta}\right)^2\nonumber\\&\times&
\left(\frac{\kappa_\phi \alpha}{1+\alpha}\right)^2
\frac{0.11 v_w^3}{0.42+v_w^2}
\frac{3.8(f/f_{\rm peak}^{\rm PT,\phi})^{2.8}}
{1+2.8(f/f_{\rm peak}^{\rm PT,\phi})^{3.8}}  ,\nonumber\\
\end{eqnarray}
where the corresponding peak frequency is expressed as~\cite{Caprini:2015zlo}
\begin{eqnarray}
f_{\rm peak}^{\rm PT,\phi}
&=&
1.65\times10^{-5}\,{\rm Hz}
\left(\frac{g}{100}\right)^{1/6}
\left(\frac{T_n}{100~{\rm GeV}}\right)\nonumber\\
&\times&
\frac{0.62}{1.8-0.1v_w+v_w^2}
\left(\frac{\beta}{\mathcal{H}}\right).
\end{eqnarray}
Here, $\mathcal{H}$ denotes the Hubble expansion rate evaluated at the nucleation temperature. The efficiency factor associated with bubble collisions is given by~\cite{Kamionkowski:1993fg}
\begin{align}
\kappa_\phi=
\frac{1}{1+0.715\alpha}
\left(
0.715 \,\alpha
+\frac{4}{27}\sqrt{\frac{3\alpha}{2}}
\right).
\end{align}

After nucleation, the expanding bubble walls interact with the surrounding thermal plasma, generating long-lasting sound waves that also source gravitational radiation. The resulting GW spectrum from sound waves is~\cite{Caprini:2015zlo,Caprini:2019egz,Guo:2020grp}
\begin{eqnarray}
\Omega_{\rm sw} h^2 &=&
2.65 \times 10^{-6}
\left(\frac{100}{g_*}\right)^{1/3}
\left(\frac{\mathcal{H}}{\beta}\right)
\left(\frac{\kappa_{\rm sw}\alpha}{1+\alpha}\right)^2
\nonumber\\&\times&v_w
\left(\frac{f}{f_{\rm peak}^{\rm PT,sw}}\right)^3
\left(
\frac{7}{4+3(f/f_{\rm peak}^{\rm PT,sw})^2}
\right)^{7/2}
\Upsilon ,\nonumber\\
\end{eqnarray}
with the peak frequency given by~\cite{Caprini:2015zlo}
\begin{eqnarray}
f_{\rm peak}^{\rm PT,sw}
&=&
1.90\times10^{-5}\,{\rm Hz}
\left(\frac{g_*}{100}\right)^{1/6}
\frac{1}{v_w}
\left(\frac{T_n}{100~{\rm GeV}}\right) 
\left(\frac{\beta}{\mathcal{H}}\right).\nonumber\\
\end{eqnarray}
The corresponding efficiency factor for sound-wave production can be approximated as~\cite{Espinosa:2010hh}
\begin{align}
\kappa_{\rm sw}=
\frac{\sqrt{\alpha}}
{0.135+\sqrt{0.98+\alpha}}.
\end{align}

The factor
$\Upsilon =
1-\frac{1}{\sqrt{1+2\tau_{\rm sw}\mathcal{H}}},$
accounts for the finite lifetime of the sound-wave period~\cite{Guo:2020grp}. The duration of the sound-wave source is estimated as $\tau_{\rm sw}\sim R/\bar{U}_f$, where the average bubble separation is
$R=(8\pi)^{1/3}v_w\beta^{-1},$
and the root-mean-square fluid velocity is
$\bar{U}_f=\sqrt{\frac{3\kappa_{\rm sw}\alpha}{4}}.$
In addition to sound waves, magnetohydrodynamic (MHD) turbulence generated in the plasma also contributes to the stochastic GW background. The corresponding spectrum is given by~\cite{Caprini:2015zlo}
\begin{eqnarray}
\Omega_{\rm turb} h^2& =&
3.35 \times 10^{-4}
\left(\frac{100}{g_*}\right)^{1/3}
\left(\frac{\mathcal{H}}{\beta}\right)
\left(\frac{\kappa_{\rm turb}\alpha}{1+\alpha}\right)^{3/2}
\nonumber\\&\times&v_w
\frac{(f/f_{\rm peak}^{\rm PT,turb})^3}
{(1+f/f_{\rm peak}^{\rm PT,turb})^{11/3}
(1+8\pi f/h_)}.
\end{eqnarray}
The peak frequency associated with the turbulence contribution is~\cite{Caprini:2015zlo}
\begin{eqnarray}
f_{\rm peak}^{\rm PT,turb}
&=&
2.88\times10^{-5}\,{\rm Hz}
\left(\frac{g_*}{100}\right)^{1/6}
\frac{1}{v_w}
\left(\frac{T_n}{100~{\rm GeV}}\right)
\left(\frac{\beta}{\mathcal{H}}\right).
\end{eqnarray}
For the turbulence efficiency factor, we adopt the commonly used approximation
\begin{align}
\kappa_{\rm turb}\simeq 0.1,\kappa_{\rm sw}.
\end{align}

Finally, the inverse Hubble time redshifted to the present epoch is expressed as
\begin{equation}
h=
1.65\times10^{-5}
\left(\frac{T_n}{100~{\rm GeV}}\right)
\left(\frac{g_*}{100}\right)^{1/6}
{\rm Hz}.
\end{equation}

To compute the phase-transition parameters $T_n$, $\beta/\mathcal{H}$, and $\alpha$, we use the finite-temperature effective potential. The bubble nucleation rate per unit volume is given by~\cite{Linde:1980tt}
\begin{equation}
\Gamma(T)\simeq T^4 \exp\left[-\frac{S_3(T)}{T}\right],
\end{equation}
where $S_3(T)$ denotes the three-dimensional Euclidean action of the critical bubble, evaluated using Findbounce~\cite{Guada:2020xnz}. The nucleation temperature $T_n$ is determined by comparing the nucleation rate with the Hubble expansion rate. The free-energy difference between the false and true vacua is defined as $\Delta V_{\rm eff}\equiv V_{\rm eff}(\phi_{\rm false},T)-V_{\rm eff}(\phi_{\rm true},T)$, from which the strength of the phase transition is characterized by the parameter $\alpha=\epsilon/\rho_{\rm rad}$. Here, the radiation energy density is $\rho_{\rm rad}=g_*\pi^2T^4/30$, while the released vacuum energy density is given by
$\epsilon=
\left[
\Delta V_{\rm eff}
-\frac{T}{4}\frac{\partial \Delta V_{\rm eff}}{\partial T}
\right]_{T=T_n}.
$
The inverse duration of the phase transition is parametrized as
$
\frac{\beta}{\mathcal{H}}
\simeq
T\frac{d}{dT}\left(\frac{S_3}{T}\right)\Bigg|_{T=T_n},
$
where $\mathcal{H}\equiv \mathcal{H}(T_n)$ is the Hubble parameter evaluated at the nucleation temperature. The bubble wall velocity $v_w$ is evaluated following the prescription given in~\cite{Lewicki:2021pgr}.

\begin{figure}[t]
    \centering
    \includegraphics[width=\linewidth]{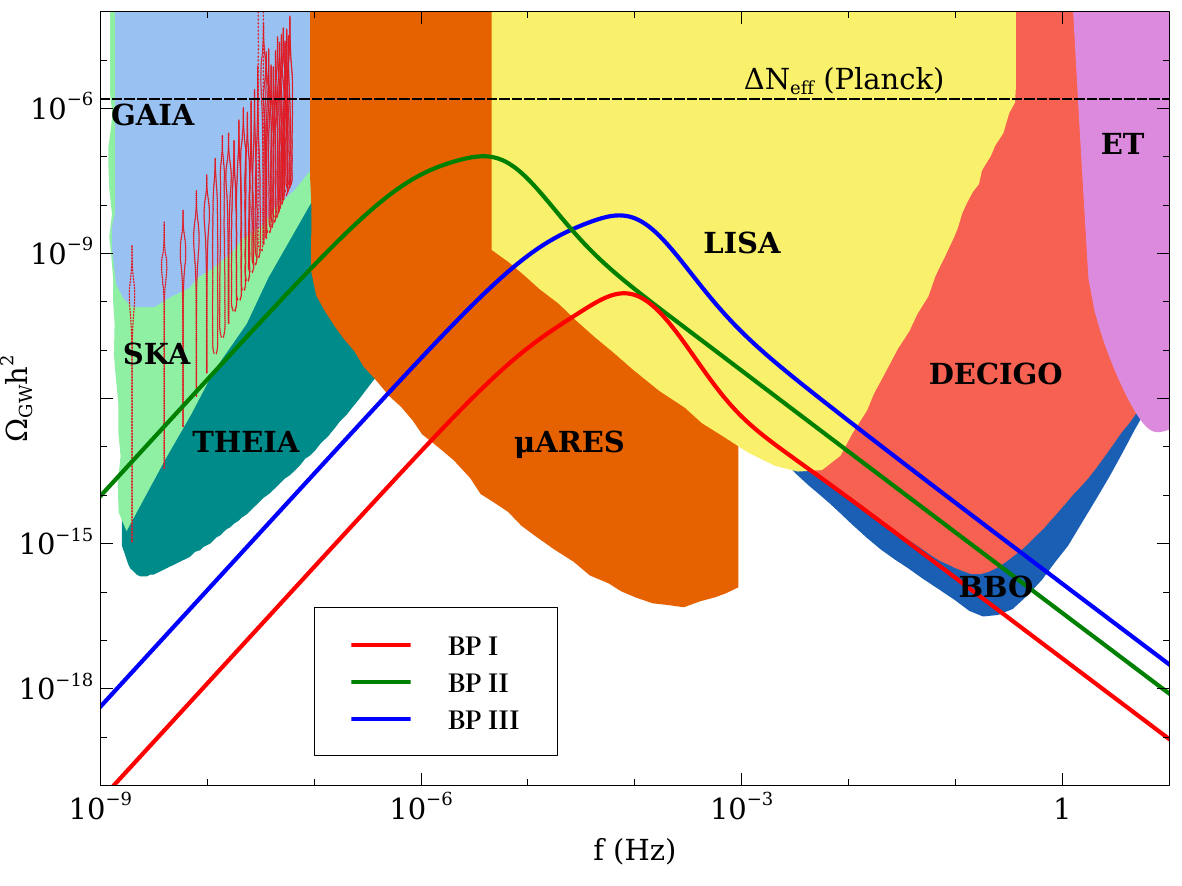}
    \caption{The FOPT-generated total gravitational wave spectra for the benchmark points listed in Table \ref{tab:BP}.}
    \label{fig:gw}
\end{figure}

Fig. \ref{fig:gw} shows the total GW spectra for the three benchmark points given in Table \ref{tab:BP}. The shaded regions correspond to the sensitivity projections of various gravitational wave experiments BBO~\cite{Yagi:2011wg}, ET~\cite{Punturo:2010zz}, LISA~\cite{LISA:2017pwj}, GAIA~\cite{Garcia-Bellido:2021zgu}, THEIA~\cite{Garcia-Bellido:2021zgu}, DECIGO~\cite{Seto:2001qf}, SKA \cite{Weltman:2018zrl}, and $\mu$ARES~\cite{Sesana:2019vho}. The dashed horizontal line denotes the upper bound on $\Delta N_{\rm eff}$ from PLANCK 2018 data \cite{Planck:2018vyg}.


\end{document}